\input{psfig.sty}

\documentstyle[11pt,aaspp4, flushrt]{article}  

\begin{document}
 
\title{Study of the Temporal Behavior of 4U 1728--34 as a Function of its
Position in the Color-Color Diagram}

\author{T. Di Salvo\altaffilmark{1,2}, 
M. M\'endez\altaffilmark{2,3}, M. van der Klis\altaffilmark{2},
E. Ford\altaffilmark{2}, N. R. Robba\altaffilmark{1}}
\altaffiltext{1}{Dipartimento di Scienze Fisiche ed Astronomiche, 
Universit\`a di Palermo, via Archirafi 36 -- 90123 Palermo, Italy.}
\authoremail{disalvo@gifco.fisica.unipa.it}
\altaffiltext{2}{Astronomical Institute "Anton Pannekoek," University of 
Amsterdam and Center for High-Energy Astrophysics,
Kruislaan 403, NL 1098 SJ Amsterdam, the Netherlands.}
\altaffiltext{3}{Facultad de Ciencias Astron\'omicas y Geof\'{\i}sicas, 
Universidad Nacional de La Plata, Paseo del Bosque S/N, 1900 La
Plata, Argentina.}

\lefthead{Temporal Behavior of 4U 1728--34}

\begin{abstract}

We study the timing properties of the
bursting atoll source 4U 1728--34 as a function of its position in the
X-ray color-color diagram.  In the island part of the color-color diagram
(corresponding to the hardest energy spectra) the power spectrum of
4U 1728--34 shows several features such as a band-limited noise component
present up to a few tens of Hz, a low frequency quasi-periodic oscillation
(LFQPO) at frequencies between 20 and 40 Hz, a peaked noise component around 
100 Hz, and one or two QPOs at kHz frequencies.  In addition to these, in the
lower banana (corresponding to softer energy spectra) we also find a very 
low frequency noise (VLFN) component below
$\sim 1$ Hz.  In the upper banana (corresponding to the softest energy 
spectra) the power spectra are dominated
by the VLFN, with a peaked noise component around 20 Hz.  We find that
the frequencies of the kHz QPOs are well correlated with the position in the 
X-ray color-color diagram.  For the frequency of the LFQPO and the break 
frequency of the broad-band noise component the relation appears more 
complex.  These frequencies both increase when
the frequency of the upper kHz QPO increases from 400 to 900 Hz, but at
this frequency a jump in the values of the parameters occurs.  We
interpret this jump in terms of the gradual appearance of a QPO at the
position of the break at high inferred mass accretion rate, while the 
previous LFQPO disappears.  Simultaneously, another kind of noise appears 
with a break frequency of $\sim 7$ Hz, similar to the NBO of Z sources.
The 100 Hz peaked noise does not seem to correlate with the position
of the source in the color-color diagram, but remains relatively constant 
in frequency.  This component may be similar to several 100~Hz QPOs observed
in black hole binaries.

\end{abstract}   


\section{Introduction}

Low mass X-ray binaries (LMXBs) are usually characterized by weak magnetic
fields.  In these systems the accretion disk can reach regions close to the 
compact object, where the matter is subject to a strong gravitational field.
This results in complex power spectra with characteristic 
frequencies, determined by the dynamical time scales close to the compact
object, up to more than a thousand Hz.  
These timing structures can be used to 
probe the physics near the compact object, and timing is a key tool to perform 
such studies.  Because several features are 
usually simultaneously present in the power spectra of LMXBs, we need to 
understand the entire timing structure in order to construct reliable theories 
for the different timing signals.  

Power spectra of LMXBs show significant changes as a function of the source
intensity and spectral hardness. Based on
their spectral and temporal behavior, LMXBs can be divided into two
classes, the so-called Z and atoll sources (Hasinger \& van der Klis
1989), whose names are derived from the tracks they trace out in an X-ray
color-color diagram.  The Z sources are thought to have higher mass accretion
rates and perhaps higher magnetic fields than the atoll sources.  These
two classes of sources show a temporal behavior that strongly depends on
their position in the color-color diagram (Hasinger \& van der Klis
1989).  In particular, atoll sources can be found in the so-called island state,
where the power spectrum is characterized by a strong band-limited noise
component, called the high frequency noise (HFN), which extends up to
tens of Hz in frequency, or in the banana state, where the power
spectrum is dominated by a power-law noise below 1 Hz, usually called
very low frequency noise (VLFN).  It is thought that these states
reflect different mass accretion rate, which increases from the island
to the banana (see van der Klis 1995).  In the island state the power
spectrum usually shows other features, such as a low frequency quasi-periodic 
oscillation (hereafter LFQPO) between 20 and 40 Hz.  This LFQPO, which may 
be similar to the horizontal branch oscillation (HBO) in Z sources, has been
variously interpreted as due to the Lense-Thirring precession of
material in the innermost disk region (Stella \& Vietri 1998), to
radial oscillations in the boundary layer between the neutron star surface 
and the accretion disk (Titarchuk \& Osherovich 1999), or in
terms of the magnetospheric beat frequency model (Alpar \& Shaham 1985;
Lamb et al.  1985; see also Psaltis et al.  1999a).

Observations with the Rossi X-Ray Timing Explorer (RXTE) have shown 
kilohertz quasi-periodic oscillations (kHz QPO) in the persistent emission 
of 20 LMXBs, with frequencies ranging from a few hundred Hz up to $1200-1300$ 
Hz (see van der Klis 2000 for a review).  Usually two kHz peaks 
are simultaneously observed.  The difference between the frequencies of
the two simultaneous kHz QPOs is, in many cases, consistent with being
constant.  During type-I X-ray bursts in some atoll sources, nearly
coherent oscillations were also detected at frequencies between 330 and
590 Hz, interpreted as the spin frequency of the neutron star
(Strohmayer et al.  1996).  In these cases the frequency separation
between the two simultaneous kHz QPOs is similar to the frequency of the
burst oscillations, or half that value (see Strohmayer, Swank, \& Zhang
1998 for a review), suggesting a beat frequency mechanism as the origin
of the kHz QPOs (Strohmayer et al.  1996; Miller, Lamb, \& Psaltis
1998).  However, the peak separation is sometimes significantly smaller
than the frequency of the burst oscillations (M\'endez, van der Klis, \&
van Paradijs 1998), and decreases as the frequency of the kHz QPOs
increases (van der Klis et al.  1997; M\'endez et al.  1998; 
M\'endez \& van der Klis 1999).  
Other models for the kHz QPOs are based on general
relativistic periastron precession at the innermost disk boundary
(Stella \& Vietri 1999), or on relaxation oscillations that occur at the
centrifugal barrier around the neutron star (Titarchuk, Lapidus, \& Muslimov
1998).

The broad-band noise component, the LFQPO and the kHz QPOs seem to be
correlated with each other and with the position of the source in the 
color-color diagram.  
The frequency of the LFQPO in Z-sources (i.e. the HBO)
is correlated with the kHz QPO frequency (e.g.  van der
Klis et al.  1997; Wijnands et al.  1998; see also Psaltis, Belloni, \&
van der Klis 1999b, and references therein), as is the LFQPO in atoll
sources (e.g.  Stella \& Vietri 1998; Ford \& van der Klis 1998; 
van Straaten et al.  2000).
The LFQPO in Z sources may differ from that in atoll sources, because  
they might be the second harmonic instead of the fundamental (cf.
Jonker et al. 1998; Wijnands \& van der Klis 1999).  
The break frequency of the broad-band
noise component in atoll sources is also correlated with the kHz QPO
frequency (Ford \& van der Klis 1998; van Straaten et al.  2000; Reig et
al.  2000) and such a component may also be present in Z-sources
(Wijnands \& van der Klis 1999).  The correlation between the LFQPO and
the kHz QPOs seems to be general, including also the $0.1-10$ Hz QPOs
observed in black hole candidates (BHCs; Psaltis et al.  1999b).  Another
general correlation seems to exist between the frequency of the break of
the broad-band noise component and the frequency of the LFQPO (Wijnands
\& van der Klis 1999).  A detailed study of the kHz QPOs in several
atoll sources revealed that there is a correlation between frequency and
position in the color-color diagram (see e.g.  M\'endez \& van der Klis 
1999; Kaaret et al. 1999; M\'endez 1999) or spectral shape (Ford et al. 1997a; 
Kaaret et al. 1998).  These correlations suggest that all these properties
depend on one parameter that could be the mass accretion rate (cf.
Hasinger \& van der Klis 1989).  On the other hand, the relation between
the frequency of the kHz QPOs and count rate is complex (Ford et al. 1997b; 
Yu et al. 1997; Zhang et al. 1998; M\'endez 1999,
and reference therein), suggesting that the X-ray flux is not a good
indicator of the mass accretion rate.

One of the most interesting sources in this context is the bursting
source 4U 1728--34.  It belongs to the class of the atoll sources, and
shows temporal variability on all time scales resulting in a complex
power spectrum.  It was observed with the EXOSAT satellite in the island
state (Hasinger \& van der Klis 1989), but a great deal of new
information about this source has recently come from RXTE observations.
4U 1728--34 shows kHz QPOs (Strohmayer, Zhang, \& Swank 1996) and burst
oscillations at 363 Hz (Strohmayer et al.  1996).  The peak separation
between the kHz QPOs in this source is always smaller than 363 Hz, and
decreases significantly at higher inferred accretion rates (M\'endez \&
van der Klis 1999).  Besides these high-frequency features, the power
spectrum of this source also shows a broad-band noise component (with a
break frequency around 10 Hz), a LFQPO (between 20 and 40 Hz;
Strohmayer et al.  1996) and a bump around 100 Hz (Ford \& van der
Klis 1998).  Both the centroid frequency of the LFQPO and the frequency
of the break of the broad-band noise seem to be correlated with the
frequency of the kHz QPOs (Ford \& van der Klis 1998).

In this paper we present new results about the correlations between
noise and QPOs in 4U 1728--34, obtained from a study of the temporal
behavior of this source as a function of the position in the
color-color diagram.  For this analysis we used all the RXTE
observations performed in 1996 and 1997, that are now publicly available
(results from these data were already published by Strohmayer et al.
1996; Strohmayer, Zhang, \& Swank 1997; Ford \& van der Klis 1998;
M\'endez \& van der Klis 1999).  We find that, while the LFQPO frequency
is correlated to both the kHz QPO and break frequencies at low inferred mass
accretion rates, at higher accretion rate there is a jump in this
correlation, probably because a new LFQPO appears at the position of the 
break, and another noise component appears in the power spectrum.

\section{Observations and Analysis}

We analyzed data of 4U 1728--34 from the public RXTE archive, using
observations performed in 1996 between February 15 and March 1, 1996 on
May 3, and 1997 between September 23 and October 1.  The log of these
observations is presented in Table 1.

From the instruments on board RXTE, we only used data from the
Proportional Counter Array (PCA), selecting intervals for which the
elevation angle of the source above the earth limb was greater
than 10 degrees.  We discarded those data ($\sim 2$\% of the total)
where 1 or 2 of the 5 PCA detectors were switched off. 
Several bursts occurred during these observations, but we excluded those
intervals ($\sim 500$ s around each burst) from our analysis.  
We used the Standard 2 mode data (16 s
time resolution) to produce light curves and the color-color diagram,
and Event data (with a time resolution of 16 $\mu$s or 122 $\mu$s) to
produce Fourier power spectra as described below.

We used the standard PCA background model 2.1b to extract background
subtracted light curves binned at 16 s in the energy bands $2.0-3.5$,
$3.5-6.4$, $6.4-9.7$, $9.7-16.0$ keV.  We defined the soft color as the
count rate ratio in the bands $3.5-6.4$ keV and $2.0-3.5$ keV, and the
hard color as the ratio of count rates in the bands $9.7-16.0$ keV and
$6.4-9.7$ keV, and produced a color-color diagram.  Because the
observations span two years, we took into account the gain changes of
the PCA between epoch 1 (until 1996 March 21) and epoch 3 (from 1996
April 15 to 1999 March 22), as well as the additional continuous gain change 
that occurs within each epoch.  As it can be seen in Table 1, 18 observations
($\sim 70\%$ of the observation time) took place in February-March 1996 during 
epoch 1, one observation occurred in May 1996 during epoch 3, and the remaining
9 observations occurred in September-October 1997, again during epoch 3.  
To correct the data for changes of the gain of the PCA, we proceeded as
described in Kuulkers et al.  (1994): we selected RXTE
observations of Crab obtained close to the dates of our observations,
and calculated X-ray colors for Crab.  Between February 1996 and
October 1997 the soft and hard colors of Crab increased by $\sim 16$\,\%
and $\sim 5$\,\% respectively, and between May 1996 and October 1997 
the same colors increased by $\sim 5$\,\% and $\sim 2$\,\%, respectively,
so we applied these (multiplicative) corrections to the colors of 
4U\,1728--34.  The resulting color-color diagram is shown in Figure 1.
The reason to use Crab colors to estimate changes in the instrumental 
response is that the spectrum of Crab is supposed to be constant, and 
therefore changes in the colors of Crab are caused by changes in the 
instrumental response.  However, the corrections calculated in this way
do not take into account the difference in spectral shape between Crab
and 4U\,1728--34.  We used spectral fits of 4U\,1728--34 and the response
matrices to test the magnitude of this effect.  The correction 
factors obtained in this way are very similar to those calculated using
Crab. They are $\sim 0.2$\,\% larger in soft color and $\sim 2$\,\% larger
in hard color than the Crab-based correction factors, less than the statistical 
errors of the colors from each 16 s interval ($\sim 3$\,\% in both soft color 
and hard color).  We preferred to use the Crab-based correction as it is 
independent of any model of either instrumental response or source spectrum.

To study the temporal variability of the source as a function of the
position in the color-color diagram we divided this diagram into 19
intervals, as shown in Figure 1.  The intervals were chosen in order to
have a good number of intervals, and good statistics in each interval.
For each interval we computed a power spectrum as follows:  we divided
the PCA light curve into segments 256 s long for which we calculated
Fourier power spectra with no selection on energy, using a Nyquist
frequency of 2048 Hz.  Because every color point spans 16 s, each power
spectrum corresponds to 16 color points.  Due to the statistical uncertainties
and to the motion of the source in the color-color diagram, a power
spectrum could be spread over more than one interval (usually 2 to 3
intervals).  For each interval, we averaged together all power spectra
that fall in that interval for more than 30\% of the time, weighing each
of them by the fraction of the time it actually spent in the interval.
We used 50\% of the time in the upper banana (intervals 18 and 19),
where the motion of the source in the color-color diagram is faster.
This selection criterion allows some overlap between the nearest
intervals.  We checked that this overlap does not affect our results.
In fact, the results we discuss in this paper do not significantly change 
if we use more restrictive selection criteria (without overlap between 
intervals), although QPO features are usually 10--20\,\% narrower in 
those cases. 
The power spectra, one for each interval selected in the color-color
diagram, are indicated with numbers from 1 to 19 starting from the
island state (upper right corner in the color-color diagram).  Each
interval contains from 10 to 134 power spectra.  The high frequency part
of each power spectrum ($1200-2048$ Hz, where neither noise nor QPOs are
known to be present) was used to estimate the Poisson noise, which was
subtracted before performing the spectral fitting.

In Figures 2a and 2b we show some representative power spectra (normalized as 
in Leahy 1983), corresponding to the intervals 4, 8, 10, 11, 12, 13, 14, and
19 of the color-color diagram of Figure 1.  Power spectra 1--11
are typical of the island state of atoll sources, while 14 and 19
correspond to the lower and the upper banana respectively (see
Hasinger \& van der Klis 1989 for definitions).  In spectra 1--11
we can distinguish several components:  the HFN, with a break
between 2 and 20 Hz, the LFQPO, between 10 and 40 Hz, a bump around
100 Hz, and one or two kHz QPOs.  We used a broken power law, 
$P(\nu) \propto \nu^{-\alpha_{1,2}}$ where $\alpha_1$ is the power-law index  
below the break and $\alpha_2$ is the power-law index above the break, 
to describe the broad-band noise, and Lorentzians to describe the LFQPO, 
the 100 Hz peaked noise, and the kHz QPOs (when present).  
The characteristic frequencies of all these features are seen to increase 
from intervals 1 to 9. In intervals 10--13 a complex
change is seen to occur in the power spectra.  One way to describe this 
change is as follows: the break of the power law becomes sharper, and
a bump is seen to occur on top of this break, which gradually turns into a 
QPO, pushing back the break to lower frequency. At the same time the original
LFQPO, while still visible, weakens.  We will discuss below how to describe
this transition in terms of fits of the power spectra.
In intervals 14 and higher, 
besides the previous components, we also see the VLFN, which we fitted using a
power law, $P(\nu) \propto \nu^{-\alpha}$, with index $\alpha \sim
1.5-2.0$.  The power spectrum of interval 19 appears much simpler than the 
others, consisting only of two noise components, namely the VLFN and the HFN.
In this case the HFN appears to have a different shape from the one we
observe in the island state, because it is peaked, i.e.  the power law 
has a positive index
before the break.  Following Hasinger \& van der Klis (1989), in this
case we modeled the HFN using a power law with $\alpha < 0$ plus an
exponential cutoff ($\nu_{\rm cut} \sim 10$ Hz).  For those intervals
where the frequencies of the kHz QPOs were high enough that the kHz QPOs
do not affect the parameters of the low frequency features and vice
versa, we fitted the kHz QPOs in a limited frequency range (500--2048 Hz) 
where no other component was present.  We then 
fitted the other components using the full frequency range and fixing
the parameters of the kHz QPOs at the values obtained in the high
frequency range.  In particular, for interval 7 and for intervals 10 to
12 we did this with upper kHz QPO, and for intervals 13 to 17 we did it
for both kHz QPOs.  For the other intervals we fitted all the components
(with all the parameters free) in the full frequency range.

The broad-band noise, the LFQPO, the 100 Hz peaked noise, and the kHz
QPOs are detected in 17 of the 19 intervals in the color-color diagram
(1 to 17), with the exception of the lower kHz peak that is absent in the
first 9 intervals.  The VLFN is detected in intervals 14 to 19.  The
broad-band noise is also detected (with a different shape) in intervals
18 and 19.  In Table 2 we report the results of these fits and in Section 3
we discuss all these results in more detail.

In Figures 2a and 2b we also plot the best fit model and the corresponding 
residuals (in units of $\sigma$) for each power spectrum.  For intervals 4 
and 5 we obtain high reduced $\chi^2$:  4.5 and 4.1 (for 131 degrees of 
freedom), respectively.  Looking at the residuals in Figure 2a, 
it is apparent that for these two intervals the main problem 
is the low frequency part of the power spectrum (below $\sim 3$ Hz, 
i.e.  the frequency of the break), where the shape of the noise 
is curved, and is badly approximated by a power law (see also Section 2.2, 
where we try an alternative model for the noise component).  Also,
the abrupt change of the slope of the broken power law around the
break frequency does not represent well the smooth change of the slope
in the data there.  Some residuals 
are also visible near the LFQPO (between 10 and 30 Hz) due to the shape 
of the QPO, that is sharper than a Lorentzian.  For intervals 15 and 16 we
obtain reduced $\chi^2$ of 2.3 and 2.6 (for 132 degrees of freedom), 
respectively.  For these two intervals the largest 
residuals appear in the region of the kHz QPOs.  In particular, the leading
wing of the lower kHz QPO is not well fitted by the model.
For all the other power spectra the reduced $\chi^2$ ranges from 1.3 to 2.0 
(for $131-139$ degrees of freedom).  These values are also rather large, but
this is not surprising given the high statistics of the power spectra and
their complexity.  Nevertheless, for these other intervals the residuals in 
Figure 2a,b are usually within 3 $\sigma$. Most of the time the  
largest residuals occur in the low frequency part of the power 
spectrum (due to the curved shape of the noise), or around the 
break (due to the sharp shape of the model at the break frequency, compared 
to the smoother shape of the data there).

From all the above we conclude that the high values of $\chi^2$ 
we obtain in some cases are mainly due to the detailed shape of 
the noise component, which is not so well approximated by a 
broken power law. However, this does not affect strongly the parameters of 
the QPOs, where the residuals are usually flat, or the 
behavior of the break frequency of the broad-band noise component,
even though a broken power law is sometimes an oversimplification of the 
complex shape of these power spectra.
Under the applied model, our estimates of the statistical errors on the fit
parameters, which use $\Delta \chi^2 = 1$, for a one-sigma single 
parameter error, therefore provide a good estimation of the total error
for the QPO parameters. For the noise parameters systematic effects probably
dominate. These will be addressed below. 

The model described above gave the best fit for most of the 
intervals.  It is also the most commonly used model in the 
literature, which allows us to compare our results to those obtained
for other sources.  However, we note that this model is not the only
possible choice. For some intervals other models can fit the data 
equally well, and more complex models can fit better. 
This non-uniqueness of the model contributes
systematic uncertainties to the parameters. In the next sections we 
try different models for the power spectra, and we discuss how the results 
can be affected by using different models.
Our conclusion will be that, independently of the model chosen to describe
the spectra, in intervals 10--12 a major change gradually occurs in the
power spectra.  In the model just described this change shows up as a jump 
in the frequency and rms amplitude of the HFN and the LFQPO (cf.  Fig. 5)
between intervals 12 and 13.  Other models can place this jump earlier
(between 9 and 10).

\subsection{Presence of a second low frequency QPO}

The results presented in the previous section were obtained by fitting
the power spectra using always the same function.  While this
facilitates the study of the power spectral parameters as a function of
the position in the color-color diagram, we must note that the power
spectra of intervals 10 to 12 are more complex than the others, and extra 
components may be required to fit them properly (cf. Fig. 2).  For those
intervals, there is an excess of power, that can be fitted by an extra
QPO, at about the break frequency of the broken power law that we used
to fit the HFN.  For instance, if we add this extra QPO to our fit function 
of interval 12, we get LFQPOs at $20.5 \pm 0.3$ Hz and $46.5 \pm 0.9$ Hz, 
while the break frequency of the HFN
decreases to $6.6 \pm 0.2$ Hz, compared to $\sim 20$ Hz when we only fit
one LFQPO to the power spectrum.  This extra LFQPO significantly
improves the fit, as $\chi^2$ decreases from 192 for 131 degrees of
freedom to 149 for 128 degrees of freedom (the F-test probability is $4 
\times 10^{-7}$).  
The addition of this extra QPO does not change the parameters of the other 
LFQPO, the 100 Hz peaked noise, and the kHz QPOs; the differences between the 
values of these parameters using these two different models are well within the 
errors.  However, the inclusion of this extra LFQPO
changes the parameters of the HFN. Both the break frequency we obtain 
for interval 12 using this model ($\sim 6.6$ Hz)
and the frequency of the extra LFQPO ($\sim 20$ Hz) are in line with
the results we obtain for interval 13.  The change in the shape of the
power spectra, that shows up in the model with only one LFQPO as a jump 
in the LFQPO frequency at interval 12--13 (to be discussed below), can be 
detected by the more complex model already in lower intervals.
The results reported in Figures 2a and 2b and in Table 2 
refer to the model with only one LFQPO.

\subsection{Fitting the broad-band noise with a Lorentzian}

For the first intervals, especially  
4 and 5, the power spectra are not flat at low frequencies, and therefore
not well fitted by a power law (cf. Fig. 2);  we then get 
better fits using a Lorentzian instead of a broken power law.  For 
interval 4, $\chi^2/d.o.f. = 587/131$ using the broken power law, and 
$461/132$ using the Lorentzian, giving a statistically significant 
improvement (although the $\chi^2$ is still large).
The Lorentzian fit becomes complicated, however, for intervals 10--12.
The sharp break there (see Fig. 2) can not be fitted with the smoother
Lorentzian.  If we fit a Lorentzian anyway, this sharp edge
appears as a QPO; two low frequency QPOs are now needed to fit
these spectra.  For interval 11, $\chi^2/d.o.f. = 241/131$ 
for the broken power law and $182/129$ using the Lorentzian plus QPO 
for the broad-band noise, again a statistically significant improvement.  
Note that an extra LFQPO also improves the fit of intervals 10--12 
when the broken power law is used to model the HFN (see Section 2.1).
For the other intervals the fits are equally good whether we use a broken
power law or a Lorentzian to represent the broad-band noise component.

The centroid frequency of the Lorentzian used to fit the broad-band
noise goes
from $0.51 \pm 0.06$ Hz to $1.0 \pm 0.1$ Hz for intervals 1 to 6, and it
is around 3.2 ($\pm 0.3$) Hz for intervals 9 to 13.  Its rms amplitude
ranges from 15 to 20\% up to interval 10; then  
it jumps down to 10\% and continues decreasing to
$\sim 3$\%.  This is the same jump that we see in the rms amplitude of
the broken power law at interval 12--13 (Fig. 5): it occurs at interval
9--10 now because in intervals 10--12 the necessary extra
QPO (centered near the break) absorbs some of the power.   
Comparing the half width at half maximum (HWHM) of the
Lorentzian (plus the centroid frequency of the Lorentzian) and $\nu_{\rm
break}$ in each interval, we see that there is a good correspondence
between these two parameters up to interval 9, with the HWHM of the
Lorentzian being a little bit larger than $\nu_{\rm break}$.
In intervals 10 to 12 $\nu_{\rm break}$ increases
(see Table 2b)
while the HWHM of the Lorentzian is almost constant around 10 Hz.
Again, this is due to the extra QPO at $\sim
20$ Hz.  For intervals 13 to 17 both the HWHM of the Lorentzian and
$\nu_{\rm break}$ are almost constant:  $\nu_{\rm break}$ is $\sim 7$ Hz
and the HWHM of the Lorentzian is $\sim 10-15$ Hz.

We conclude that the choice of either of these two models to fit the
broad-band noise component does not affect the behavior of the
parameters that we obtain, 
except for the width of the noise after interval 12.
In fact, the parameters of all the
other components that we obtain using a Lorentzian instead of the broken
power law for the broad-band noise are consistent (usually within $1 \sigma$;
in a few cases the parameters of the 
LFQPO are within $3 \sigma$) with the values 
already discussed, and show the same behavior when plotted against the 
frequency of the upper kHz QPO.  

The addition of an extra LFQPO near 20 Hz in intervals 10--12, 
whether we use a Lorentzian for the HFN or a broken power 
law, modifies the width of the noise (break frequency or Lorentzian HWHM) 
and its rms amplitude, which jump to lower values ($\sim 7$ Hz and 
$\sim 7\%$, respectively).
These results are in line with the 
parameters of the HFN and LFQPO in interval 13 and higher, indicating that
a gradual change in the shape of the power spectra
begins in intervals 10--12.

Finally, we also tried to fit these spectra using zero-centered
Lorentzians for the noise and the bump at 100 Hz, as in Olive et al.
(1998) for the source 1E 1724--3045.  We did not
obtain good fits, the principal reason being that the 100 Hz peaked
noise can not be well fitted by a zero-centered Lorentzian for most of
the intervals.

\section{Results} 
  
We now describe the results from the fit with the model described in Section
2. We will note differences with the other fit function where appropriate
in Section 4.

In Figure 3a we plot the frequencies of both kHz QPOs (upper panel) 
and their rms amplitudes (lower panel) as a function of
the interval number in the color-color diagram. We also show the kHz QPO
rms amplitudes as a function of the corresponding kHz QPO frequency
in Figure 3b.
While in the power spectra of intervals 10
to 17 we see both kHz QPOs, in the power spectra
of intervals 1 to 9 (at low inferred mass accretion rate) we only see
one of them, which in principle could be either the upper or the lower 
peak.  However, when we plot the rms amplitude of the kHz QPOs vs.  the
position in the color-color diagram 
for all the intervals in which we detect kHz
QPOs (Fig.  3a, lower panel), the rms amplitude of the single kHz QPO 
connects smoothly with the rms amplitude of the upper kHz QPO, whereas
at the same time the rms amplitude of the lower kHz QPO is significantly
lower. The same behavior is also visible in Figure 3b, where we plot the
rms amplitude of each kHz QPO as a function of the corresponding frequency.  
We therefore conclude that when we only see one of the kHz QPO,
it is always the upper peak (see also Ford \& van der Klis 1998; M\'endez
\& van der Klis 1999).

The frequencies of the kHz QPO range from $\sim 400$ to $\sim
1200$ Hz for the upper peak, and from $\sim 500$ to $\sim 900$ Hz for
the lower peak.  The frequencies of both QPOs are well correlated to the
position of the source in the color-color diagram, although the frequency
of the upper kHz QPO seems to jump from $\sim 500$ to $\sim 700$ Hz
between intervals 6 and 7.  This jump does not seem to be present  
in the plot of kHz QPO frequency versus position of the source in the 
color-color diagram (indicated with $S_{\rm a}$) shown in 
M\'endez \& van der Klis (1999; Fig. 4).  However, this  
is because the jump is hidden by the intrinsic fluctuations of the kHz QPO 
frequency in the plot of M\'endez \& van der Klis (1999).  
If we average together points with similar values of $S_{\rm a}$ in 
their plot, we find that the average frequency of the kHz QPO remains more 
or less constant at $\sim 500$ Hz for 
$S_{\rm a} < 1.4$, similar to what we observe for intervals 3--6,
and then jumps to $\sim 700$ Hz for $S_{\rm a} \simeq 1.5$.

For both kHz QPOs the rms amplitude ranges from 3\% to $\sim 12\%$, but while 
the rms amplitude of the upper peak generally decreases with frequency, the 
rms amplitude of the lower peak increases with frequency (except for the
last point, which is significantly lower than the previous one).  
The width of the upper peak
generally decreases with frequency.  This implies higher values of the
quality factor $Q=\nu/\Delta \nu$ (which ranges from $\sim 1.5$ to $\sim
9.5$), and therefore higher coherence of the signal, at higher
frequency.  For the lower peak no clear trend is evident.  In this case
Q ranges between 3.5 and 7.4 and is $\sim 15$ in the last interval in
which the lower peak is detected.  We note that in each interval we
averaged many power spectra, and therefore the measured width of the kHz
QPOs could be affected by changes in their centroid frequencies within
the interval.  
In particular, we can estimate the effect of the frequency variations on the 
width of these features, considering the relation found for the frequency of 
the kHz QPOs versus interval number, and knowing that each power spectrum 
can be spread over at most three intervals. For interval numbers higher 
than 7, we find that the frequency of the kHz QPOs varies by $\sim 150$ Hz 
across 3 intervals. This is comparable with the measured widths of the kHz 
QPOs in interval 7 and higher where, therefore, the widths of the kHz QPOs
are likely considerably affected by this broadening effect. For interval 
numbers lower than 7, the frequency of the kHz QPOs varies by $\sim 50$ Hz 
across 3 intervals. This is much smaller than the measured width of the kHz
QPO (more than 200 Hz FWHM) in these intervals, indicating that the kHz QPOs  
may be intrinsically broader at low frequencies.

The 100 Hz peaked noise is a broad feature, with a FWHM that
ranges from 100 to more than 200 Hz.  Its frequency
(Fig.  4, upper panel) seems to fluctuate around 130 Hz, without any
clear correlation with the kHz QPO frequencies, although at higher values of
the upper kHz QPO frequency ($\nu_{\rm high} > 900$ Hz) it seems to increase 
with $\nu_{\rm high}$.  Figure 4 (lower panel) shows the rms 
amplitude of the 100 Hz bump as a function of the frequency
of the upper kHz QPO.  For $\nu_{\rm high} < 
900$ Hz the rms amplitude of the 100 Hz bump stays constant at
$\sim 10$\%, and for $\nu_{\rm high} > 900$ Hz it decreases to $\sim 4$\%.
It seems to be anti-correlated with the corresponding centroid frequency,
at least at higher kHz QPO frequencies.

The broken power-law index below the break ($\alpha_1$) ranges from 
$-0.09 \pm 0.04$ to $0.07 \pm 0.02$, and the index above the break 
($\alpha_2$) is almost constant
around 1.5 (the typical error is 0.1), except for intervals 10 to 12
where it ranges between 2 and 2.5.  This is because the shape of the 
power spectrum at the frequency of the break becomes sharp in these 
intervals, and probably another
component begins to appear here (see Section 2.1).  In Figure 5 (upper
panel) we plot the break frequency of the broad-band noise component
($\nu_{\rm break}$) vs.  the frequency of the upper kHz QPO.  The break
frequency increases from $\sim 2$ to $\sim 25$ Hz when $\nu_{\rm high}$
increases from $\sim 400$ Hz to 900 Hz.  Beyond this frequency,
$\nu_{\rm break}$ jumps down to 5 Hz and then remains constant around 7
Hz.  The rms amplitude of the broad-band noise component (measured in
the whole frequency range; Fig.  5, lower panel)
generally decreases as the source moves in the color-color diagram from 
the island to the banana,
as expected.  In particular, it remains at $\sim 15\%$ up to
$\nu_{\rm high} = 800$ Hz where it starts decreasing; at $\nu_{\rm high}
= 900$ Hz, it jumps down to 5\%, and continues decreasing down to $\sim
3$\% as $\nu_{\rm high}$ increases.

Similar jumps at the same value of $\nu_{\rm high} = 900$ Hz are also
visible in the frequency and rms amplitude of the LFQPO (Fig.  5, upper
and lower panel respectively).  For $\nu_{\rm high} < 900$ Hz the LFQPO
frequency increases from 8 to 46 Hz as $\nu_{\rm high}$ increases.  At
$\nu_{\rm high} \sim 900$ Hz it jumps down to $\sim 20$ Hz, it then
starts increasing again with $\nu_{\rm high}$, and finally seems to saturate.
The rms amplitude of the LFQPO initially decreases from 15\% to 3\% as
$\nu_{\rm high}$ increases up to 900 Hz; at that point the rms amplitude
of the LFQPO jumps up to 10\% and then decreases again from 10\% to 5\%
as $\nu_{\rm high}$ increases.

Table 2b shows the parameters of the VLFN as a function of the interval
number in the color-color diagram.  The rms amplitude of the VLFN increases 
(from $\sim 1$\% to $\sim 5\%$) with increasing inferred mass accretion rate,
while the index of the power law is almost constant at 1.7.  For
intervals 18 and 19 we fitted the broad-band noise using a power law
(with a positive index around 1.5) with an exponential cutoff at a
frequency of $\sim 10$ Hz.  The rms amplitude of this component is
around 2\%.

\section{Discussion}

The results illustrated in the previous section show that both the
frequency and the rms amplitude of the kHz QPOs and X-ray colors are 
all well correlated with each other.
It is thought that the position in the color-color 
diagram reflects different mass accretion rate (see e.g.  Priedhorsky 
et al. 1986; Hasinger \& van der Klis 1989; Hasinger et al. 1990; van der Klis
1994; Kaaret et al. 1998).  
Moreover, various models of the kHz QPOs 
associate their frequencies with the Keplerian frequency at the inner 
rim of the accretion disk (Miller et al. 1998; Titarchuk et al. 1998; 
Stella \& Vietri 1999). The inner radius of the disk is thought to 
decrease with increasing mass accretion rate (Miller et al. 1998), suggesting 
that higher kHz QPO frequencies correspond to higher accretion rate.
These considerations suggest that the mass accretion rate 
determines both the position of the source in the color-color diagram and 
the related timing behavior (cf. Hasinger \& van der Klis 1989). 

In the first 9 power spectra we only see
one kHz QPO.  In principle we can not say, in this case, if it is the
lower or upper peak.  Ford \& van der Klis (1998) tentatively identified
this QPO as the upper peak.  This identification is in line with the
result of M\'endez \& van der Klis (1999) that, in general, the single
kHz QPO shows the same correlation with the inferred mass accretion rate 
as the upper peak.  
This identification is strengthened by our results presented here that there 
is a smooth trend of rms amplitude with QPO frequency if the single QPOs are 
identified with the upper of the two kHz QPOs (see Fig. 3b).
This confirmation is important because the correlation
of the LFQPO frequency with the frequency of
the kHz QPOs lines up with that of other sources only if we make the
identification as we do here (see e.g.  Psaltis et al.  1999b).

We find that the parameters of the LFQPO and the broad-band noise are
correlated with the frequency of the kHz QPOs in a complex way.  As we
can see in Figure 5, the frequencies of the LFQPO and of the break are
correlated with the frequency of the upper kHz QPO up to $\nu_{\rm high}
\sim 900$ Hz, which corresponds to interval 12, just before the 
appearance of the VLFN (see Table 2b).  Beyond this
frequency there is a jump in both the break and the LFQPO frequencies,
confirming the trend already pointed out by Ford \& van der Klis (1998).
After the jump the frequency of the LFQPO seems to increase and then
saturate at $\sim 40$ Hz while, except for the first point, $\nu_{\rm
break}$ seems to be constant around 7 Hz.
If for intervals 10--12 we use a model with an extra LFQPO
(see Sections 2.1 and 2.2), the break
of the HFN jumps to lower values, $\sim 6.6$ Hz. These values are
similar to the break frequency measured in interval 13 and higher,
indicating that 
the change in power spectral shape already begins in intervals 10--12
(corresponding to $\nu_{\rm high} \sim 850-900$ Hz).
We also note that the frequency of the extra LFQPO ($\sim 20$ Hz) is similar 
to that of the LFQPO in intervals 13, suggesting that the LFQPO 
in intervals 13 and higher is the extra LFQPO present in 10--12, 
while the LFQPO at higher frequencies (present up to interval 12) disappears.

As mentioned in Section 2, looking carefully at the power 
spectra (Fig.  2), something is seen to
happen starting at interval 10, namely an excess appears on top of the
break at a frequency that is around half that of the LFQPO, while the
LFQPO itself is fading as can be seen from its rms amplitude (Fig.  5, lower
panel).  In interval 12 two QPOs, one at 20 Hz (where we put the break
frequency of the broken power law) and one at 46 Hz, could be present
(see Sec. 2.1).
So, the jump in the parameters of the LFQPO at $\nu_{\rm
high} \sim 900$ Hz can be due to the fact that beyond this frequency 
we measure the parameters of what seems
to be another LFQPO, which begins to appear at the
frequency of the break already in intervals 10 to 12.  At the same point,
there is a jump in the parameters of the
HFN (see Fig.  5 at $\nu_{\rm high} \sim 900$ Hz).  This change can
be identified in the power spectra: in interval 13 the HFN fits what 
appears to be another noise component characterized by a different 
behavior and a drastically lower break frequency ($\sim 7$ Hz).  
Some evidence of the emergence of this noise is 
already visible in the earlier power spectra (see Sec. 2.1).  
Jumps are also visible in 
the rms amplitudes of the noise component and of the LFQPO at $\nu_{\rm
high}=900$ Hz (Fig.  5, lower panel).  We note that at the same
frequency similar variations (more gradual in this case) are also
present in the rms amplitudes of the upper kHz QPO and the 100 Hz peaked
noise, which both decrease from about 10\% to $\sim 4$\% (Fig.  3b and
4).

A possible interpretation of the behavior of the LFQPO and broken
power-law break frequencies is that the noise component at low inferred
accretion rates turns into a QPO at higher accretion rates, while what
we call the LFQPO at low accretion rates gradually disappears at higher
accretion rates.  In Figure 5 (top) we can see that the centroid frequency of
the LFQPO after the jump is in line with that of the break, as expected
in this interpretation.  A continuity is also visible between the rms
amplitudes of the broken power law before the jump and the LFQPO after
the jump (Fig. 5, bottom).  We observe that the alternative models for 
intervals 10 to 12, containing an extra LFQPO, are also consistent 
with this interpretation, although the continuity between the rms amplitudes 
of the broken power law and the LFQPO is not observed in this case,
because some of the power is absorbed by the extra LFQPO.
After the jump the value of $\nu_{\rm break}$ seems to be
constant around 7 Hz, which is in the same range of frequencies of the normal
branch oscillations (NBOs) observed in the Z sources. 
 
No QPO like the
NBO are usually observed in atoll sources.  For this reason these QPOs
were associated with high accretion rates.  Recently Wijnands et al.
(1999) found that 4U 1820--30 shows a 7 Hz QPO at its highest mass accretion 
rate levels (see also Dotani et al. 1989).
Going from the lower to the upper banana, in this source the broad-band 
noise seems to evolve into a broad peaked noise and then into a QPO around 
7 Hz (Wijnands et al.  1999).  This noise near 7 Hz seems quite similar to 
the noise after the jump shown by 4U 1728--34.  

So, in 4U 1728--34 when mass
accretion rate increases we see a noise component near 20 Hz turn into a
QPO, while a new noise component appears at 7 Hz.  This component at 7
Hz is similar to the one in 4U 1820--30, which in that source turns into
a QPO when mass accretion rate increases further.
We note that the noise we find after the jump in 4U 1728--34 seems to
become peaked, i.e.  before interval 13 the index $\alpha_1$ of the broken
power law (below the break) is $\sim 0$, but after the jump at interval 
13, $\alpha_1$ becomes negative (see Table 2b).  A similar behavior was also 
observed in 4U 0614+09 (van Straaten et al.  2000). 

In order to compare our results on 4U 1728--34 with similar results from
other sources,
we plotted in Figure 6 the frequencies of the break of the broad-band noise 
component (filled triangles) and of the LFQPO (open diamonds) on a 
logarithmic scale as a function of the lower kHz QPO frequency 
$\nu_{\rm low}$.  For the first 9 intervals we calculated the frequency
of the lower kHz QPO by subtracting an assumed constant peak separation of
363 Hz from the frequency of the upper kHz QPO.
The frequency separation between the twin kHz QPOs in 4U 1728--34 is known
to be constant for $\nu_{\rm low}$ between 600 and 800 Hz (M\'endez \& van der
Klis 1999), and we are here extrapolating this result down to lower frequencies.
We caution that it is unknown whether this peak separation remains constant
down to the lowest kHz QPO frequencies.  For instance, the relativistic 
precession model for the kHz QPOs (Stella \& Vietri 1999) predicts that the peak
separation should decrease at low frequencies (probably below $\nu_{\rm high}
\sim 500-600$ Hz, depending on the mass of the compact object).  This might
explain the deviation of the points at the lowest frequencies from the 
observed correlations in Figure 6. A smaller peak separation
would move these points towards higher values of $\nu_{\rm low}$, closer
to the observed correlations.

Using the interpretation described above, we can distinguish three low
frequency features, that are plotted in the diagram shown in Figure 6.  
The first one is
the LFQPO at low accretion rate, that is correlated with the lower kHz 
QPO frequency (and therefore with the inferred
mass accretion rate).  As already known (Ford \& van der Klis 1998), it
follows a power law of index $\sim 0.95$ (the solid line drawn in Figure
6), the same as observed for the HBOs of the Z sources (Psaltis et al.
1999a,b. Note that line in Fig. 2 of Psaltis et al. 1999b has an
index of 1.05).  The second
feature is the break of the broad-band noise which turns into a LFQPO at high
accretion rate.  This feature also seems to be correlated with the frequency
of the lower kHz QPO.  It follows a power law with index $\sim 1.5$ (the
dashed line in Fig. 6).
The third feature is the break
frequency at high accretion rate, which we identify with the NBO of the
Z sources and the 7 Hz QPO of 4U 1820--30.

Variabilities such as QPOs and broad noise components, have been
observed in different kinds of accreting X-ray sources, like Z sources,
atoll sources and BHCs.  Recently some authors
(Wijnands \& van der Klis 1999; Psaltis et al.  1999b) have 
studied correlations between these components in order to understand if
they could be explained by a single mechanism. Psaltis et al.  (1999b)
propose that two kinds of variability can be identified, one
similar to the HBO of the Z sources and one similar to the kHz QPO, both
in neutron star and in BHC systems, with different coherences and range
of frequencies.  The data used by Psaltis et al.  (1999b) are also shown
in Figure 6 (gray points).  
In the case of the BHC systems a $0.1-10$ Hz QPO is
often accompanied by a $\sim 1-200$ Hz broad noise.  Psaltis et al.  (1999b)
suggest to identify the first QPO 
with the HBO and the noise with the ``kHz'' QPO.  When these
two kinds of variabilities are plotted one against the other (PBK
diagram; Psaltis et al.  1999b) a strong correlation appears that seems to 
extend the correlation observed for the Z sources over three orders of 
magnitude.  In addition, the PBK
diagram shows two more branches below this main branch, a second branch
described by the BHCs XTE 1550--564 and GRO 1655--40, and the atoll
source 4U 1608--52, and a third branch occupied by the NBO of the Z
sources.  If we compare our results on 4U 1728--34 with the plot shown in 
Psaltis et al.  (1999b), we can see that our dashed line 
describing the break frequency at low mass accretion rate and the LFQPO at 
high mass accretion rate coincides with the second branch in PBK, and that 
$\nu_{\rm break}$ after the jump coincides with the third branch
(see Fig.  6).

Another general correlation seems to exist between the frequencies of
the LFQPO and the break for Z sources, atoll sources and BHCs, that is
discussed in Wijnands \& van der Klis (1999).  These data are plotted
in Figure 7 (gray points).  In order to compare our
results with this correlation (hereafter WK correlation)
we plotted in Figure 7 the LFQPO frequency as function of $\nu_{\rm break}$ 
for 4U 1728--34 (filled triangles).  We see that all the points 
before the jump follow the WK correlation, whereas the four points 
after the jump lie at $\nu_{\rm break} \sim 7$ Hz and 
$\nu_{\rm LFQPO} \sim 40$ Hz in Figure 7, 
slightly above the main relation, i.e. in the part of the diagram occupied 
by the Z sources.
These correspond to higher accretion rates and follow a similar, but 
slightly shifted correlation, perhaps because the noise component is 
different in these sources (Wijnands \& van der Klis 1999).
The position of these points in the
diagram also seems to suggest that, in 4U 1728--34
after the jump, another kind of variability is present, characterized by 
a different behavior.  Note that with our proposed
identification of $\nu_{\rm break}$ at low mass accretion rate with the
second branch in the PBK diagram, we can now understand the WK
correlation as the correlation between the main branch and the second
branch in the PBK diagram. By plotting in the PBK diagram the break frequency
vs. the lower kHz QPO frequency for other atoll sources this second branch
will probably become further populated.

To summarize, we find that at low inferred mass accretion rate the
frequency of the LFQPO and $\nu_{\rm break}$ are well correlated with the
frequency of both kHz QPOs (and therefore with the inferred mass accretion 
rate).
These frequencies describe two branches in the PBK diagram that are
related through the WK relation.  At high inferred mass accretion rate
the previous LFQPO disappears and the broad-band noise changes into a
QPO with a centroid frequency that corresponds to the break frequency of
the previous noise.  In the PBK diagram this new LFQPO falls in the same branch 
as the break frequency before the change.  A new broad-band noise appears at 
these high accretion rates, characterized by a break
frequency around 7 Hz.  This could correspond to the NBO observed in Z
sources, and also in the atoll source 4U 1820--30, at high accretion
rate.  All these results seem to indicate that there is a tight relation
between noise and QPOs.  In fact there is evidence that noise could
change into a QPO and vice versa.  Moreover, their characteristic
frequencies show a precise dependence on inferred accretion rate, which shows
up as the three branches in the PBK diagram.

The jump in the frequency of the LFQPO can be understood in the framework
of the magnetospheric beat frequency model assuming that the
neutron star magnetic field is approximately but not exactly
dipolar (e.g., a dipole offset from the center of the star).
In this case, at lower accretion rates, when the magnetospheric
radius is far from the neutron star, the gas in the disk feels 
primarily the dipolar field. This is symmetric, and the LFQPO
frequency will be double the difference between the orbital frequency
at the magnetospheric radius and the spin frequency.  At
higher accretion rates, when the magnetospheric radius
moves closer to the stellar surface, the gas in the disk feels primarily 
the offset, asymmetric field and the beat frequency will be
the difference between the orbital and the spin frequency. 

However, a phenomenology where LFQPO can mutate into broad-band noise and vice
versa is probably more easily accommodated by models in which the LFQPO
arises due to disk fluctuations (e.g., Titarchuk \& Osherovich 1999;
Psaltis \& Norman 1999) than by those in which they are produced by
orbital motion at the magnetospheric radius (Psaltis et al. 1999a)
or Lense-Thirring precession of free-orbiting self-luminous blobs at the 
inner disk whose orbital radius is the upper kHz QPO frequency (Stella \&
Vietri 1998), particularly if the kHz QPO maintain a high Q while at the
same time the LFQPO turns into broad noise. We point out, however, that
disk fluctuation models do appear to be able to provide variability at or
near the frequencies predicted by such orbital models (Alpar \& Yilmaz
1997; Psaltis \& Norman 2000).

The 100 Hz broad bump seems {\em not} to follow these general
correlations, but instead to remain relatively constant in frequency
at low inferred mass accretion rate. At higher accretion rate the 
frequency increases, while the rms amplitude decreases.
We note that this component might be similar to the QPOs between
67 and 300 Hz that have been found in several BHCs
(Morgan, Remillard, \& Greiner 1997; Remillard et al.  1999a; 
Remillard et al.  1999b; Homan et al. 2000).  A similar QPO near 100 
Hz is also seen in the atoll source 4U 0614+09 (van Straaten et al.  2000).

\acknowledgments
The authors thank F. Lamb and L. Burderi for useful and 
stimulating discussions.
This work was supported by the Italian Space Agency (ASI), by the
Ministero della Ricerca Scientifica e Tecnologica (MURST), by the
Netherlands Research School for Astronomy (NOVA), the Netherlands
Organization for Scientific Research (NWO) under contract number
614-51-002 and the NWO Spinoza grant 08-0 to E.  P.  J.  van den Heuvel.
MM is a fellow of the Consejo Nacional de Investigaciones
Cient\'{\i}ficas y T\'ecnicas de la Rep\'ublica Argentina.  This
research has made use of data obtained through the High Energy
Astrophysics Science Archive Research Center Online Service, provided by
the NASA/Goddard Space Flight Center.


\clearpage

\begin{table}[th]
\begin{center}
\scriptsize
\caption{RXTE observations of 4U 1728--34 used in our analysis.
The count rate corresponds to 5 PCA units and is corrected for background.}
\label{tab1}
\begin{tabular}{c|c|c} 
\tableline \tableline
Start Time  & Total Time & Averaged Rate   \\ 
 (UTC)      & (ksec)     &  (c/s)           \\ 
\tableline
15/02/'96 11:51 & 25     &  1863            \\ 
15/02/'96 18:59 & 18     &  1948              \\
16/02/'96 00:03 & 20     &  2060             \\
16/02/'96 06:26 & 15     &  2172             \\
16/02/'96 15:49 & 21     &  2160               \\
16/02/'96 22:15 & 10     &  2199              \\
18/02/'96 11:10 & 25     &  1609             \\
18/02/'96 19:15 & 15     &  1516            \\
22/02/'96 11:33 & 25     &  996              \\
22/02/'96 19:12 & 15     &  997             \\
23/02/'96 21:16 & 30     &  978              \\
24/02/'96 05:17 & 5      &  1090            \\
24/02/'96 17:51 & 25     &  1022           \\
25/02/'96 01:15 & 12     &  1034           \\
25/02/'96 20:51 & 25     &  1080            \\
26/02/'96 04:02 & 15     &  1093           \\
29/02/'96 23:12 & 0.8    &  1200            \\
01/03/'96 00:13 & 20     &  1193           \\
03/05/'96 14:03 & 2.1    &  1126           \\
23/09/'97 23:50 & 1.2    &  2190           \\
24/09/'97 09:14 & 25     &  2582            \\
24/09/'97 17:15 & 4      &  2422           \\
26/09/'97 12:30 & 25     &  1775            \\
27/09/'97 09:19 & 25     &  1506           \\
30/09/'97 04:34 & 15     &  1799          \\
30/09/'97 10:57 & 8      &  1793           \\
01/10/'97 06:09 & 25     &  1659             \\
01/10/'97 14:10 & 3.5    &  1683           \\
\tableline
\end{tabular}
\end{center}
\end{table}

\clearpage 

\begin{table}[th]
\begin{center}
\tiny
\tablenum{2a}
\caption{Fitted parameters of the kHz QPOs and the peaked noise
at 100 Hz.}
\label{tab2}
\begin{tabular}{ccccccccccc}
\tableline 
 
         &       & \multicolumn{3}{c}{Upper kHz QPO} & \multicolumn{3}{c}{Lower kHz QPO} & \multicolumn{3}{c}{100 Hz peaked noise} \\ 
Interval & Rate\tablenotemark{a}& Frequency& FWHM & rms & Frequency & FWHM  &rms\tablenotemark{b}& Frequency & FWHM  & rms\tablenotemark{b} \\ 
Number   & (c/s) &  (Hz)     & (Hz)  & (\%)&  (Hz)     & (Hz)  & (\%) &  (Hz)     & (Hz)  & (\%) \\ 
\tableline
1      & 1315 & 387$\pm$18 & 230$\pm$73 &   9.6$\pm$2.1  & - & - & - & 64$\pm$37 & 350$\pm$110 & 16.3$\pm$3.1 \\  
2      & 1285 & 397$\pm$14 & 270$\pm$37 & 10.57$\pm$0.97 & - & - & - & 110$\pm$20 & 269$\pm$61 & 12.5$\pm$2.0 \\
3      & 1209 & 466$\pm$21 & 295$\pm$52 & 12.0$\pm$1.2 & - & - & - & 163$\pm$15 & 205$\pm$61 & 10.4$\pm$1.6 \\
4      & 1156 & 497.8$\pm$5.3 & 282$\pm$17 & 12.80$\pm$0.34 & - & - & - & 154.7$\pm$3.6 & 164$\pm$14 & 10.51$\pm$0.44 \\ 
5      & 1155 & 498.7$\pm$5.9 & 287$\pm$18 & 12.44$\pm$0.35 & - & - & - & 151.5$\pm$4.0 & 183$\pm$16 & 10.79$\pm$0.46 \\
6      & 1151 & 517.3$\pm$8.1 & 219$\pm$26 & 11.10$\pm$0.59 & - & - & - & 169.8$\pm$9.6 & 220$\pm$33 & 11.22$\pm$0.86 \\
7      & 1375 & 694.5$\pm$9.3 & 206$\pm$33 & 13.31$\pm$0.68 & - & - & - & 153$\pm$11 & 118$\pm$47 & 8.9$\pm$1.8 \\
8      & 1510 & 729.0$\pm$3.5 & 162.6$\pm$8.5 & 11.77$\pm$0.26 & - & - & - & 106$\pm$21 & 211$\pm$38 & 10.7$\pm$1.8 \\
9      & 1605 & 789.6$\pm$2.5 & 143.4$\pm$5.7 & 11.52$\pm$0.19 & - & - & - & 72$\pm$24 & 239$\pm$27 & 12.5$\pm$2.0 \\
10     & 1722 & 846.9$\pm$2.0 & 133.0$\pm$6.0 & 10.93$\pm$0.19 & 511$\pm$17 & 98$\pm$48 & 3.12$\pm$0.54 & 94.6$\pm$9.0 & 190$\pm$15 & 12.14$\pm$0.90 \\
11     & 1739 & 872.9$\pm$1.6 & 131.1$\pm$4.3 & 10.54$\pm$0.13 & 559$\pm$11 & 88$\pm$32 & 2.98$\pm$0.35 & 91.4$\pm$5.4 & 167.5$\pm$9.2 & 11.39$\pm$0.54 \\
12     & 1732 & 905.2$\pm$2.6 & 134.5$\pm$7.4 & 10.12$\pm$0.22 & 599$\pm$14 & 139$\pm$30 & 4.60$\pm$0.39 & 103.9$\pm$5.4 & 149$\pm$12 & 10.76$\pm$0.57 \\
13     & 1865 & 948.8$\pm$3.8 & 158$\pm$12 & 8.30$\pm$0.26 & 673$\pm$10 & 191$\pm$33 & 5.82$\pm$0.41 & 121.9$\pm$3.0 & 98$\pm$10 & 6.86$\pm$0.33 \\
14     & 1871 & 1054.7$\pm$8.7 & 166$\pm$27 & 5.62$\pm$0.33 & 750.9$\pm$3.7 & 121.8$\pm$9.3 & 7.04$\pm$0.22 & 142.1$\pm$5.4 & 87$\pm$14 & 4.54$\pm$0.34 \\
15     & 1907 & 1105.8$\pm$6.8 & 148$\pm$19 & 4.79$\pm$0.22 & 773.0$\pm$1.8 & 103.8$\pm$5.1 & 7.16$\pm$0.13 & 159.1$\pm$5.4 & 94$\pm$16 & 3.85$\pm$0.29 \\
16     & 1861 & 1129$\pm$13 & 117$\pm$22 & 3.38$\pm$0.27 & 816.0$\pm$4.0 & 166.0$\pm$6.6 & 7.82$\pm$0.13 & 146.5$\pm$8.7 & 102$\pm$16 & 3.68$\pm$0.27 \\
17     & 1939 & 1158$\pm$18 & 129$\pm$56 & 3.68$\pm$0.54 & 876.1$\pm$2.6 & 59.5$\pm$7.2 & 5.59$\pm$0.24 & 210$\pm$32 & 180$\pm$74 & 3.96$\pm$0.74 \\
18     & 2500 & - & - & - & - & - & - & - & - & - \\
19     & 2600 & - & - & - & - & - & - & - & - & - \\
\tableline
\end{tabular}
\end{center}
\tablenotetext{}{Errors correspond to $\Delta \chi^2 = 1$.}
\tablenotetext{a}{PCA count rate not corrected for the background.
The background average count rate of the PCA is 132 c/s.}

\vspace{1cm}

\begin{center}
\tiny
\tablenum{2b}
\caption{Fitted parameters of the broad-band noise, the low-frequency
QPO and the very low frequency noise. }
\label{tab3}
\begin{tabular}{ccccccccccc}  
\tableline  
       &             \multicolumn{4}{c}{HFN}                  &               \multicolumn{3}{c}{LFQPO}       &            \multicolumn{2}{c}{VLFN} &  \\ 
Interval & $\alpha_1$ &   break or cutoff & $\alpha_2$ & rms & Frequency     & FWHM             & rms\tablenotemark{a} & $\alpha$ &rms\tablenotemark{b}&  $\chi^2 (d.o.f.)$  \\ 
Number  & $(\times 10^{-2})$ &  (Hz)   &               & (\%)    &   (Hz)        & (Hz)             & (\%)          &        & (\%) &  \\ 
\tableline
1 & $-1.5$$\pm$1.5 & 1.392$\pm$0.038 & 1.55$\pm$0.14 & 13.7$\pm$1.1 & 8.08$\pm$0.43 & 15.43$\pm$0.79 & 17.00$\pm$0.95 & - & -        & 241 (131) \\
2 & $-2.4$$\pm$1.2 & 1.364$\pm$0.055 & 1.35$\pm$0.11 & 15.4$\pm$1.7 & 8.63$\pm$0.30 & 15.19$\pm$0.74 & 15.93$\pm$0.85 & - & -        & 273 (131) \\
3 & 2.6$\pm$1.1    & 1.891$\pm$0.039 & 1.348$\pm$0.061 & 14.74$\pm$0.90 & 9.04$\pm$0.58 & 24.10$\pm$0.95 & 17.15$\pm$0.78 & - & -        & 270 (131) \\
4 & 2.91$\pm$0.57  & 2.804$\pm$0.031 & 1.543$\pm$0.058 & 14.07$\pm$0.50 & 13.37$\pm$0.36 & 26.01$\pm$0.69 & 16.02$\pm$0.46 & - & -        & 587 (131) \\
5 & 3.10$\pm$0.55  & 2.817$\pm$0.029 & 1.467$\pm$0.043 & 14.63$\pm$0.46 & 13.84$\pm$0.32 & 25.88$\pm$0.65 & 15.48$\pm$0.40 & - & -        & 538 (131) \\
6 & 1.2$\pm$1.0    & 2.978$\pm$0.058 & 1.422$\pm$0.082 & 15.00$\pm$0.90 & 15.10$\pm$0.62 &  27.1$\pm$1.4  & 15.02$\pm$0.75 & - & -        & 267 (131) \\
7 & 6.8$\pm$1.7    &  8.94$\pm$0.58  & 1.314$\pm$0.090 & 18.4$\pm$1.3 & 26.17$\pm$0.85 &  12.8$\pm$3.9  &  6.2$\pm$1.0  & - & -        & 219 (134)\tablenotemark{c} \\
8 & 4.77$\pm$0.87  &  10.39$\pm$0.21 & 1.385$\pm$0.089 & 15.8$\pm$1.0 & 28.99$\pm$0.57 &  20.0$\pm$2.7  & 7.04$\pm$0.59 & - & -        & 282 (131) \\
9 & 2.27$\pm$0.77  & 13.90$\pm$0.22  & 1.511$\pm$0.098 & 14.94$\pm$0.92 & 35.58$\pm$0.46 &  12.0$\pm$1.6  & 4.44$\pm$0.33 & - & -        & 254 (131) \\
10& 2.21$\pm$0.77  & 19.05$\pm$0.29  & 2.05$\pm$0.21 & 12.23$\pm$0.57 & 41.63$\pm$0.80 &  16.3$\pm$2.7  & 4.20$\pm$0.51 & - & -        & 275 (131)\tablenotemark{c} \\
11& 1.54$\pm$0.63  & 21.08$\pm$0.22  & 2.22$\pm$0.16 & 11.62$\pm$0.35 & 45.16$\pm$0.75 &  12.7$\pm$2.5  & 3.16$\pm$0.25 & - & -        & 241 (131)\tablenotemark{c} \\
12& 1.78$\pm$0.93  & 23.63$\pm$0.41  & 2.62$\pm$0.31 & 10.86$\pm$0.42 & 46.52$\pm$0.93 &  13.5$\pm$3.9  & 3.22$\pm$0.60 & - & -        & 192 (131)\tablenotemark{c} \\
13& $-8.2$$\pm$4.3 &  5.06$\pm$0.32  & 1.42$\pm$0.18 & 5.26$\pm$0.67 & 22.22$\pm$0.76 &  39.2$\pm$2.0  & 10.61$\pm$0.38 & - & -        & 181 (134)\tablenotemark{d} \\
14& $-9.2$$\pm$4.7 &  6.60$\pm$0.40  & 1.29$\pm$0.10 & 5.56$\pm$0.45 & 36.99$\pm$0.60 &  28.3$\pm$2.0  &  6.54$\pm$0.24 & 1.64$\pm$0.19 & 1.05$\pm$0.12   & 203 (132)\tablenotemark{d} \\
15& $-2.4$$\pm$3.9 &  6.91$\pm$0.39  & 1.299$\pm$0.094 & 4.72$\pm$0.32 & 40.44$\pm$0.33 &  22.0$\pm$1.1  &  5.55$\pm$0.13 & 1.68$\pm$0.11 & 1.284$\pm$0.091 & 305 (132)\tablenotemark{d} \\
16& $-4.9$$\pm$5.5 &  7.06$\pm$0.61  & 1.34$\pm$0.11 & 4.03$\pm$0.29 & 41.77$\pm$0.36 &  21.3$\pm$1.3  &  5.16$\pm$0.14 & 1.784$\pm$0.080 & 1.82$\pm$0.11   & 348 (132)\tablenotemark{d} \\
17& $-9$$\pm$19    &   6.5$\pm$1.3   & 1.31$\pm$0.57 & 3.11$\pm$0.97 & 41.0$\pm$1.1  &  25.8$\pm$3.9  &  4.80$\pm$0.36 & 1.718$\pm$0.087 & 3.39$\pm$0.31   & 191 (132)\tablenotemark{d} \\
18& -              &  14.6$\pm$4.4   & $-1.38$$\pm$0.51& 2.60$\pm$0.16 &  - &  -               &  -               & 1.712$\pm$0.033 & 5.66$\pm$0.26   & 179 (139) \\
19& -              &  9.38$\pm$3.0   & $-1.54$$\pm$0.61& 2.16$\pm$0.16 &  - &  -               &  -               & 1.602$\pm$0.028 & 4.73$\pm$0.17   & 184 (139) \\   
\tableline
\end{tabular}
\end{center}
\tablenotetext{}{Errors correspond to $\Delta \chi^2 = 1$.}
\tablenotetext{b}{rms amplitude calculated between 0.001 and 1 Hz.}
\tablenotetext{c}{For this interval the reported $\chi^2$ is calculated
in the whole frequency range fixing the parameters of the upper kHz peak
at the values found in a limited frequency range.}
\tablenotetext{d}{For this interval the reported $\chi^2$ is calculated
in the whole frequency range fixing the parameters of both kHz QPOs
at the values found in a limited frequency range.}
\end{table}

\clearpage

\begin{figure}[h]
\centerline
{\psfig
{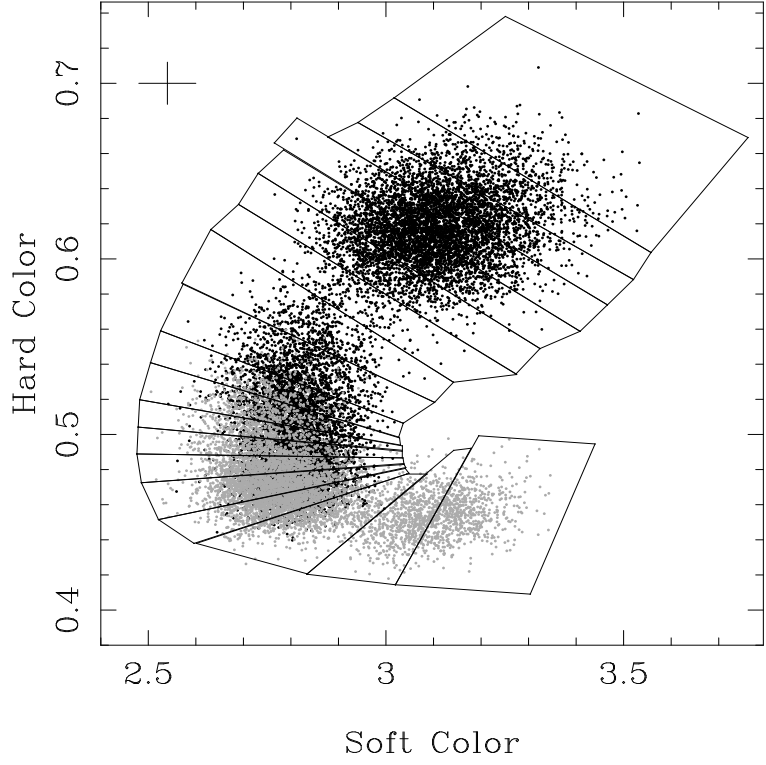}}
\caption{Color-color diagram of 4U 1728--34.  The soft and hard 
colors are defined as the ratio of the count rate in the bands 
$3.5-6.4$ keV/$2.0-3.5$ keV, and $9.7-16$ keV/$6.4-9.7$ keV, 
respectively.  Each point represents 16 s of data.  Black points represent
epoch 1 data (until 1996 March 21), and gray points represent epoch 3 data
(from 1996 April 15 to 1999 March 22).  A typical error bar is
shown at the top left corner of the diagram.  The boxes in the color-color 
diagram indicate the intervals we used to select the power spectra.}
\label{fig1}
\end{figure}

\begin{figure}[t!]
\centerline
{\psfig
{figure=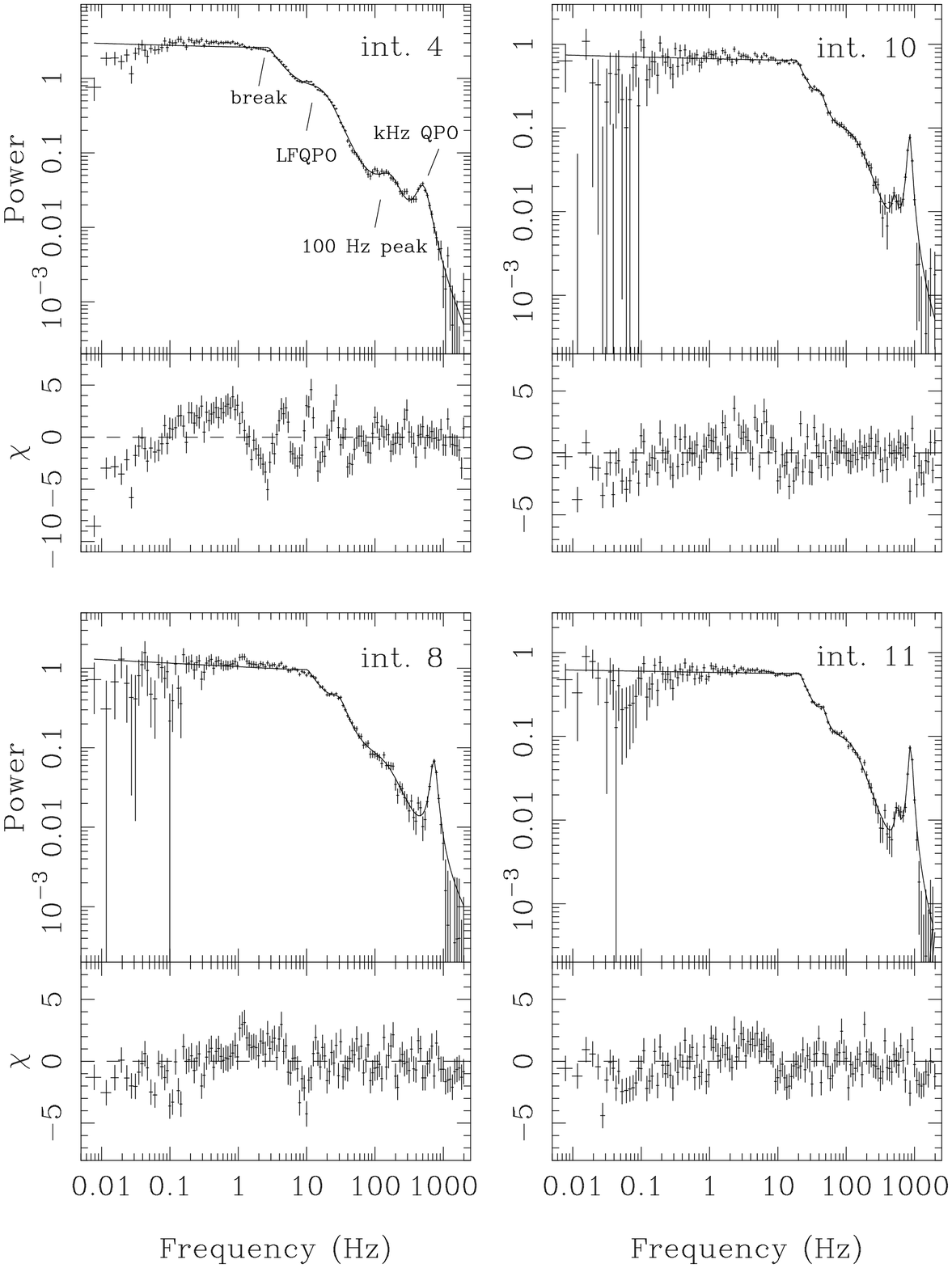,height=16.0cm,width=14.0cm}}
Fig. 2a.   \hspace{0.2cm}
Representative Leahy normalized power spectra of 4U 1728--34 
corresponding to different intervals in the color-color diagram of Figure 1. 
Interval numbers are indicated.  For each interval the power spectrum and the 
best fit model (including always at most one LFQPO, see text) are shown in the
upper panel, and the corresponding residuals (in units of $\sigma$) are shown
in the lower panel.
\label{fig2a}
\end{figure}

\begin{figure}[t!]
\centerline
{\psfig
{figure=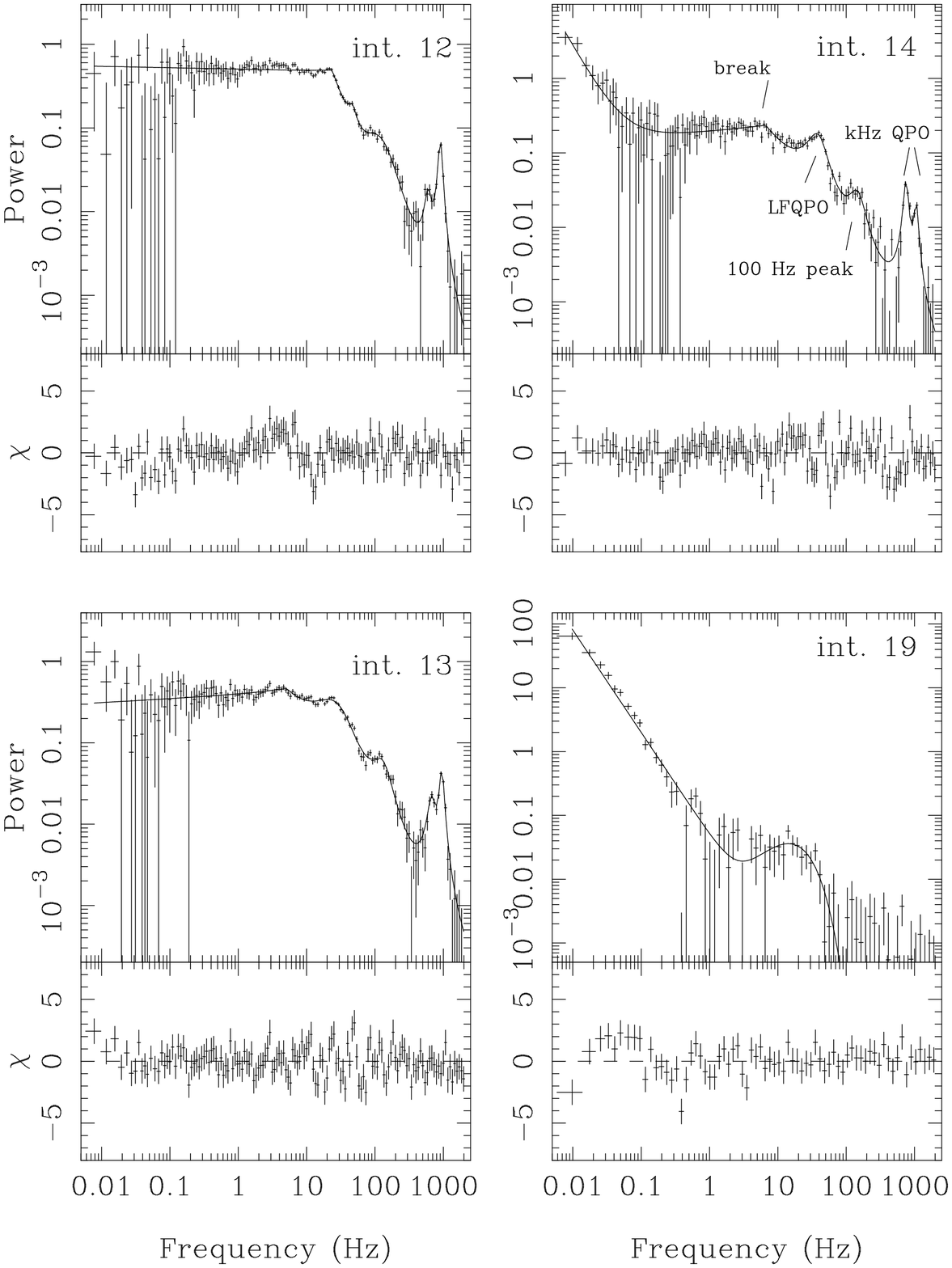,height=16.0cm,width=14.0cm}}
Fig. 2b.   \hspace{0.2cm}
Same as Figure 2a. 
\label{fig2b}
\end{figure}

\begin{figure}[t!]
\centerline
{\psfig
{figure=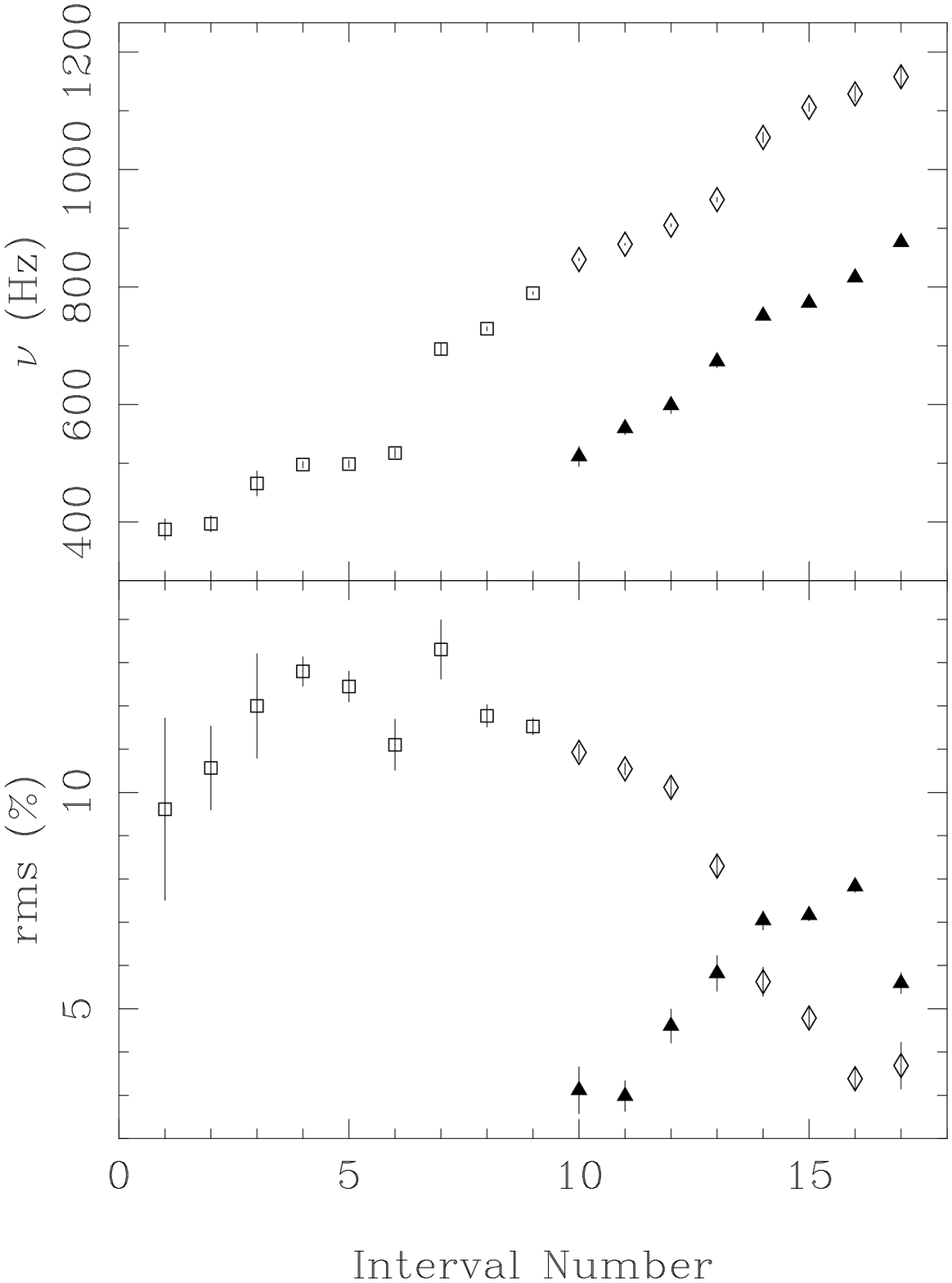,height=10.0cm,width=10.0cm}}
Fig. 3a.   \hspace{0.2cm}
Frequencies (upper panel) and rms amplitudes (lower panel) of 
the kHz QPOs as a function of the interval number in the color-color diagram.  
Different symbols are used for the lower peak (filled triangles) and the upper 
peak (open diamonds).  Open squares represent intervals where we only detected 
one of the kHz QPOs.
\label{fig3a}

\centerline
{\psfig
{figure=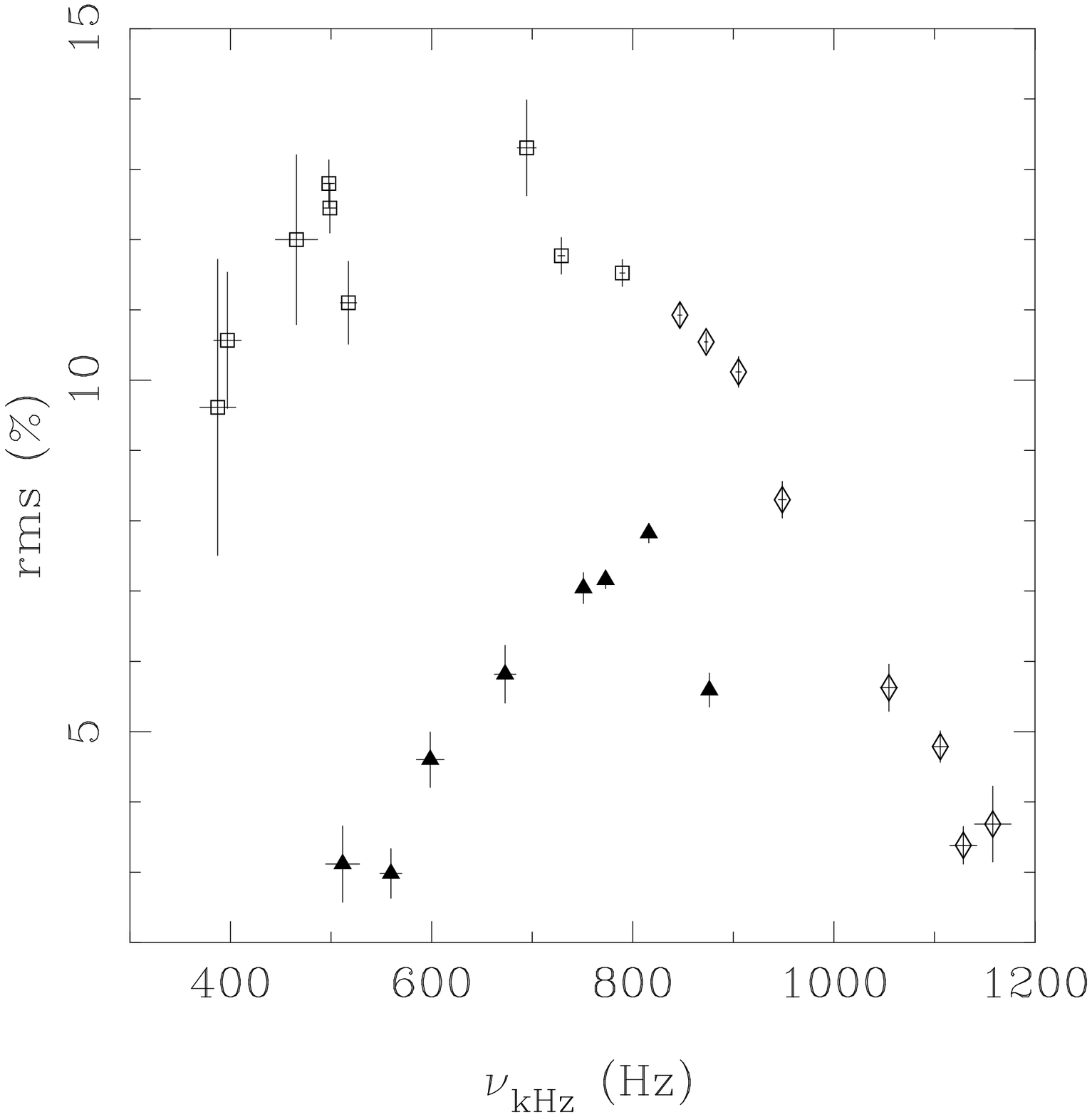,height=10.0cm,width=10.0cm}}
\vskip -0.8cm
Fig. 3b.   \hspace{0.2cm}
Rms amplitudes of the kHz QPOs as a function of the 
corresponding kHz QPO frequency. Symbols are as described in Figure 3a. 
\label{fig3b}
\end{figure}

\begin{figure}[t!]
\centerline
{\psfig
{figure=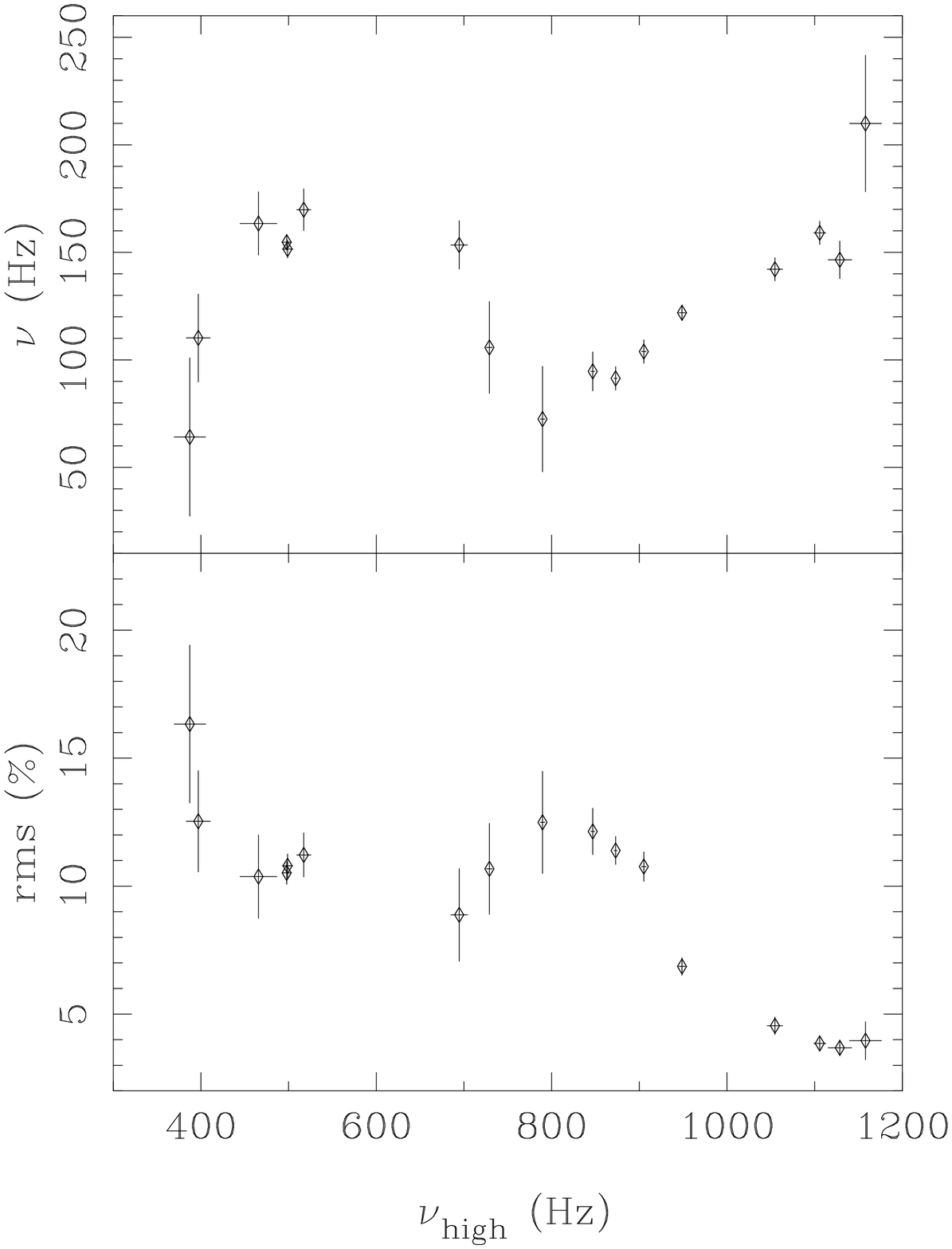,height=9.0cm,width=10.0cm}}
Fig. 4.   \hspace{0.2cm}
Frequency (upper panel) and rms amplitude (lower panel)
of the 100 Hz peaked noise as a function of the frequency of the upper kHz 
QPO.
\label{fig4}

\centerline
{\psfig
{figure=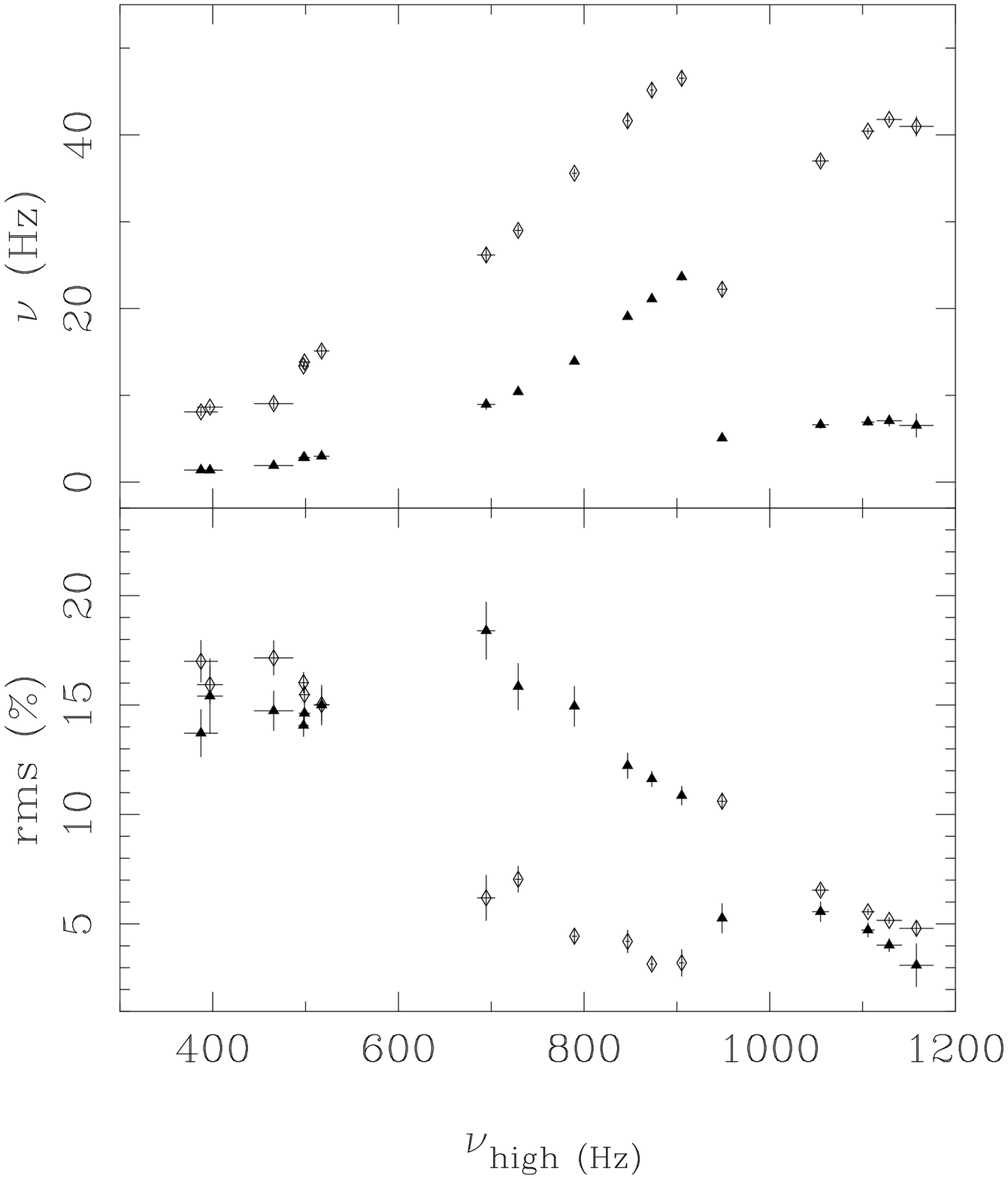,height=10.0cm,width=10.0cm}}
Fig. 5.   \hspace{0.2cm}
Parameters of the HFN and LFQPO as a function of the 
frequency of the upper kHz QPO.  Different symbols are used for the HFN 
(filled triangles) and LFQPO (open diamonds). 
Upper panel: break frequency of the HFN and centroid 
frequency of the LFQPO.  Lower panel: rms amplitudes of the HFN (calculated
from 0 Hz to infinity) and of the LFQPO. 
\label{fig5}
\end{figure}

\begin{figure}[t!]
\centerline
{\psfig
{figure=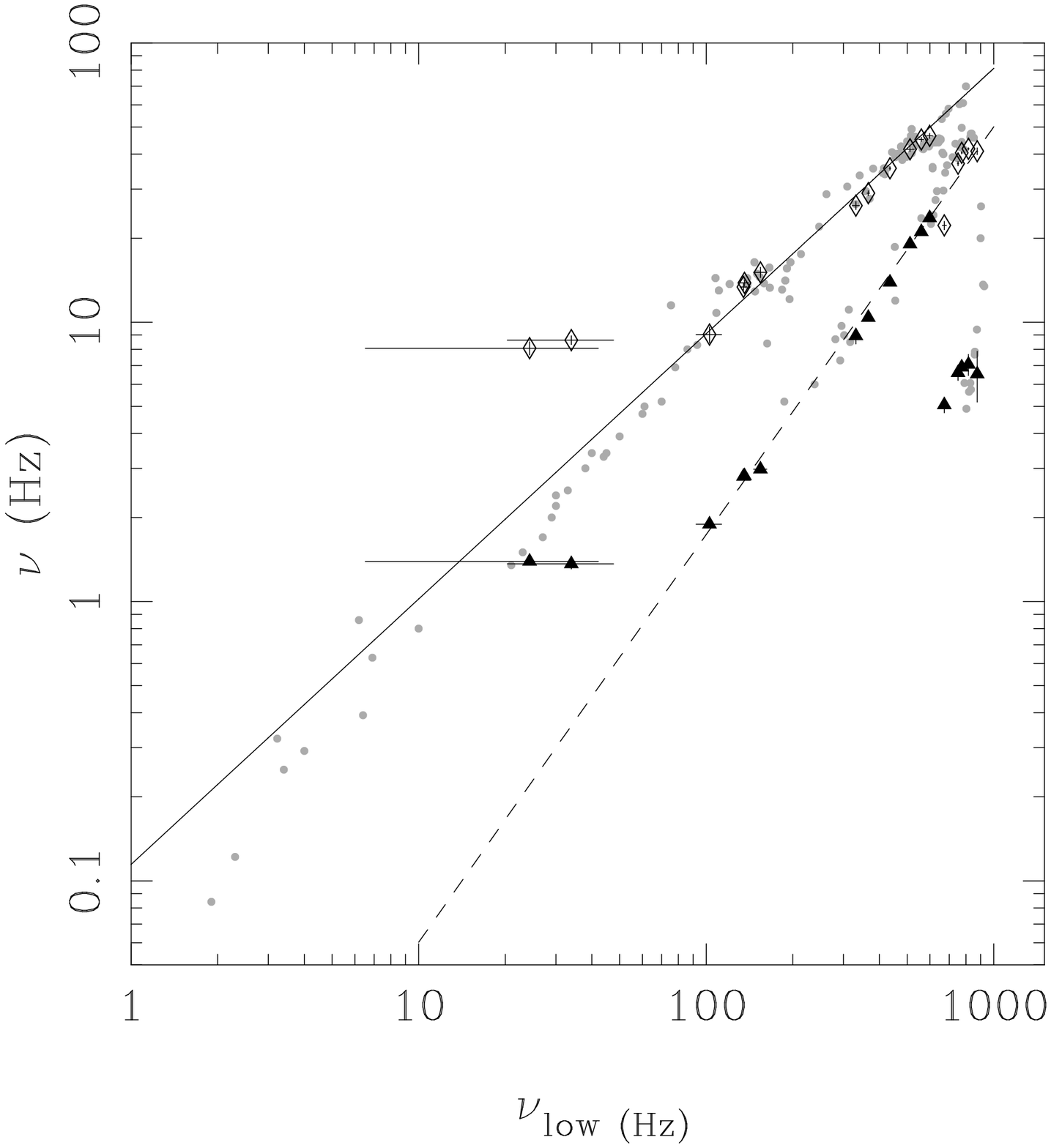,height=16.0cm,width=14.0cm}}
Fig. 6.   \hspace{0.2cm}
Frequencies of the LFQPO (open diamonds) and of the break 
(filled triangles) as a function of the frequency of the lower kHz QPO, 
$\nu_{\rm low}$.  In the first 9 intervals $\nu_{\rm low}$
is calculated by subtracting the assumed constant peak separation of 363 Hz 
from the frequency of the upper kHz QPO that has been detected.
For comparison, data for several LMXBs (including Z sources, atoll sources
and black hole candidates, from Psaltis et al.  1999b) are also shown (gray
points).  Correlations between these frequencies are visible:
the solid line is a power law of index 0.95 (Psaltis et al.  1999b) and the 
dashed line is a power law of index 1.5.
\label{fig6}
\end{figure}

\begin{figure}[t!]
\centerline
{\psfig
{figure=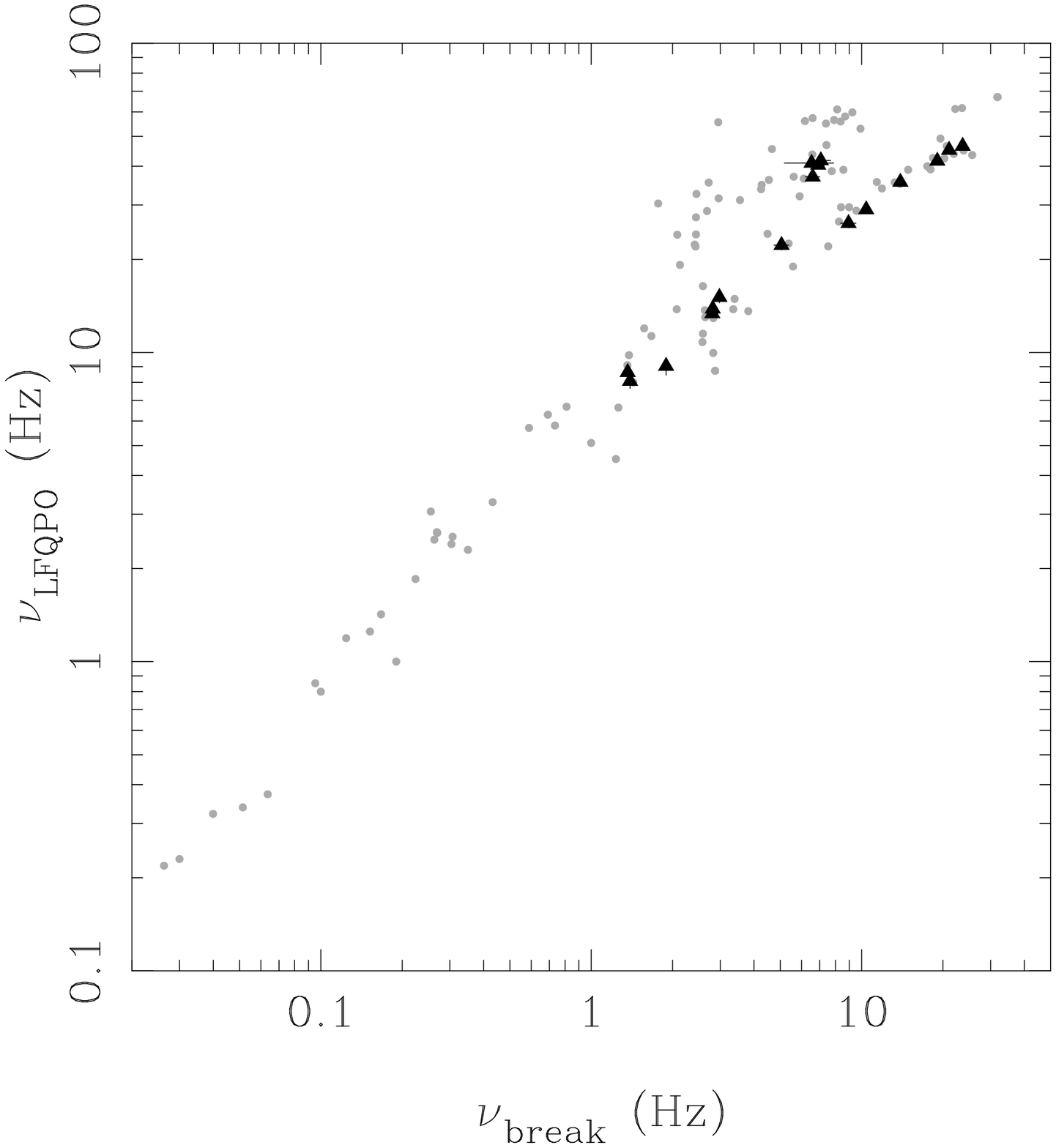,height=16.0cm,width=14.0cm}}
Fig. 7.   \hspace{0.2cm}
Centroid frequency of the LFQPO as a function of the break 
frequency of the HFN (filled triangles).  For comparison, data for several 
LMXBs (including Z sources, atoll sources and black hole candidates, from 
Wijnands \& van der Klis 1999) are also shown (gray points).
These frequencies follow the WK correlation, except 
for four points at $\nu_{\rm break} \sim 7$ Hz and $\nu_{\rm LFQPO} \sim
40$ Hz, corresponding to frequencies of the upper kHz QPO higher than 900 Hz.
\label{fig7}
\end{figure}

\end{document}